\documentclass[a4paper]{jpconf}
\usepackage{graphicx}

\usepackage{amsmath,url}
\usepackage{amssymb,amsfonts,color,helvet,graphics}
\usepackage[TABTOPCAP,tight]{subfigure}
\usepackage{epsfig}
\usepackage{pstricks}
\usepackage{setspace}
\usepackage{multirow}

\newcommand{\Jsq}{\hat{J}^2}

\newcommand{\Nmax}{N_{\mathrm{max}}}

\definecolor{title}{rgb}{.8,0,0}
\definecolor{frametitle}{rgb}{.2,.2,.7}
\definecolor{institution}{rgb}{.1,.3,.8}
\definecolor{shade}{rgb}{0,.1,.1}
\definecolor{goodgreen}{rgb}{.15,.6,.3}
\definecolor{zinn}{rgb}{.92,.76,.68}
\definecolor{beige}{rgb}{.96,.96,.86}
\definecolor{purple}{rgb}{.5,0,.5}
\definecolor{gray}{rgb}{.6,.6,.6}
\definecolor{peach}{rgb}{1,.9,.7}
\definecolor{pr}{rgb}{.5,0,0}
\definecolor{pg}{rgb}{.56,1,.44}
\definecolor{pb}{rgb}{0,0,.56}
\definecolor{po}{rgb}{.99,.67,0}
\newcommand{\ignore}[1]{}
\newcommand{\codes}[1]{{\textsf{\small #1}}}    

\begin{document}
\title{Advancing Nuclear Physics Through TOPS Solvers and Tools}

\author{
E~Ng$^1$,   
J~Sarich$^2$,  
S~M~Wild$^2$, 
T~Munson$^2$, 
H~Aktulga$^1$,  
C~Yang$^1$, 
P~Maris$^3$,  
J~P~Vary$^3$, 
N~Schunck$^{4}$, 
M~G~Bertolli$^5$, 
M~Kortelainen$^{5,6}$,  
W~Nazarewicz$^{5,6}$,  
T~Papenbrock$^{5,6}$, 
M~V~Stoitsov$^{5,6}$}

\address{\scriptsize $^1$ Computational Research Division, Lawrence Berkeley National Laboratory, Berkeley, CA
94720, USA}
\address{\scriptsize $^2$ Mathematics and Computer Science Division, Argonne National Laboratory, Argonne, IL 60439, USA}
\address{\scriptsize $^3$ Department of Physics, Iowa State University, Ames, IA 50011, USA}
\address{\scriptsize $^4$ Physics Division, Lawrence Livermore National Laboratory, Livermore, CA 94551, USA}
\address{\scriptsize $^5$ Department of Physics and Astronomy, University of Tennessee, Knoxville, TN 37996, USA}
\address{\scriptsize $^6$ Physics Division, Oak Ridge National Laboratory, Oak Ridge, TN 37831, USA}

\ead{egng@lbl.gov}

\newcommand{\gras}[1]{\boldsymbol{#1}}

\begin{abstract}
At the heart of many scientific applications is the solution of
algebraic systems, such as linear systems of equations, eigenvalue
problems, and optimization problems, to name a few.  TOPS,
which stands for Towards Optimal Petascale Simulations, is a SciDAC
applied math center focused on the development of solvers for
tackling these algebraic systems, as well as the deployment of such
technologies in large-scale scientific applications of
interest to the U.S. Department of Energy.  In this paper, we highlight
some of the solver technologies we have developed in optimization
and matrix computations.  We also describe some accomplishments
achieved using these technologies in UNEDF, a SciDAC application 
project on nuclear physics.
\end{abstract}


\section{Introduction}

Over the last couple of decades, simulation science has become as
important as theoretical and experimental science.  The success of
simulation science hinges on the ability to perform the calculations
efficiently.  The inner most kernel in these calculations is often
the solution of algebraic systems, including, but not limited
to, systems of linear and nonlinear equations, eigenvalue problems,
optimization problems, and sensitivity analysis.  TOPS, which stands
for {\em Towards Optimal Petascale Simulations\/}, is a multi-institutional
SciDAC applied math center that focuses on the development of solvers
for tackling these algebraic systems, as well as the deployment of
such technologies in large-scale scientific applications, particularly
those of interest to the U.S. Department of Energy.

In this paper, we highlight two specific areas of TOPS: eigenvalue
calculations and optimization.  In particular, we highlight some
accomplishments we have achieved in collaboration with
computational physicists in UNEDF.  The goal of the UNEDF SciDAC
application project~\cite{UNEDF}
is to obtain a comprehensive understanding of
nuclei and their reactions based on the most accurate knowledge
of the strong nuclear interaction.  Eigenvalue calculations come
up in the solution of the nuclear Schr\"odinger
equation~\cite{scidac,sc2008}.  
The eigenvalues and the eigenvectors correspond to the energy states and
wave functions.  Numerical optimization techniques are needed in
building the next generation of nuclear energy functionals, which
will provide nuclear physicists better tools for predicting the
properties and behavior of atomic nuclei over the entire nuclear
table.

%


\section{Eigenvalue Calculations}
In nuclear configuration interaction calculation,
it is sometimes necessary to investigate, among others, nuclear level 
densities as a function of the total angular momentum $J$ and 
excitation energy, and to evaluate scattering amplitudes as a function 
of $J$~\cite{shirokov}.
We will refer to this as a {\em total-J calculation\/} in this paper.
In this type of calculation, we are 
interested in computing a relatively large number of states
with a prescribed $J$ value. 


One brute-force approach to a total-J calculation is to simply compute 
a large number of eigenvalues and wave functions of a nuclear many-body 
Hamiltonian, for example in an M-scheme basis (good angular momentum 
projection along the z-axis), and select, among these wave functions,
the ones that have a prescribed $J$ value. This approach is appropriate 
when the number of desired energy states and wave functions is small 
(e.g., ten to twenty states).  When that is not true, or when certain 
properties of a nucleus pertaining to a fixed $J$ are to be calculated, 
the brute-force approach 
may require computing a very large number of wave functions, and the 
computational cost for performing this type of calculation may be 
prohibitively high. Furthermore, even if we can afford to perform this 
type of calculation, this may not be an efficient use of resources 
because we compute a large number of wave functions only to throw away 
most of them because they do not have the desired $J$ value.

We have developed an alternative approach where we construct an invariant 
subspace $Z$ that contains all wave functions associated with a fixed $J$ 
value in advance and project the nuclear many-body Hamiltonian into this 
subspace to produce a projected Hamiltonian with the minimum dimension 
consistent with that chosen $J$. A sparse matrix diagonalization procedure 
\cite{arpa:95,arpack,sore:92} is then applied to this projected Hamiltonian 
to obtain the desired energy states and their corresponding wave functions. 

To construct $Z$, we need to work with the total angular momentum 
square operator $\Jsq$ and compute the null space of $\Jsq - \lambda I$, 
where $\lambda = J(J+1)$ is a known eigenvalue of $\Jsq$.

When the many-body basis states associated with the configuration 
space are properly ordered and grouped, $\Jsq$ becomes block diagonal: 
$
\Jsq = \mbox{diag}(\Jsq_1, \Jsq_2, ..., \Jsq_{n_g}).
$  
Therefore, the task of computing the desired null space of $\Jsq-\lambda I$ 
reduces to that of computing the desired null spaces of $\Jsq_i - \lambda I$, 
for $i = 1,2,...,n_g$.  

However, because
the dimensions of the $\Jsq_i$'s vary over a wide range 
(e.g., from 1 to more than 36,000 for $^{12}$C, $\mbox{N}_{\mbox{max}}=6$),
it is difficult to 
maintain a good load balance in the null space calculation. 
Here, $\mbox{N}_{\mbox{max}}$
is a parameter limiting the total number of oscillator quanta
allowed in the many-body states.
%

We developed a multi-level task and data distribution scheme to 
achieve optimal parallel performance in the null space calculation
by
\begin{small}
\begin{enumerate} 
\item Limiting the granularity of the parallelism; that is, we try to
      divide the overall task into many small tasks of limited sizes
      so that good load balance arises from distributing these small
      tasks evenly among different processors.
\item Limiting the communication overhead incurred in the null space
      calculation so that the overall time of the computation can be
      minimized.
\end{enumerate}
\end{small}

To achieve these inherently conflicting goals, we classified $\Jsq_i$ 
blocks into small, medium and large groups based on the estimated 
computational loads associated with computing the desired null space of 
$\Jsq_i -\lambda I$, and the estimated ratio of communication volume to 
floating point operations count.

The small $\Jsq_i$ blocks are distributed among all processors based
on their computational load by a greedy algorithm.  The null spaces
of these matrices are computed by a sequential \codes{LAPACK} rank-revealing QR 
subroutine. No communication is involved in these calculations. 
Each one of the medium-sized $\Jsq_i$ blocks is 
assigned to a subgroup of processors by the same greedy algorithm.
The null space calculation for such a block is parallelized 
among processors within the same subgroup, which will incur some
communication overhead. 
Finally, the desired null space calculation for a large $\Jsq_i$ block
is carried out in parallel on all processors.

We implemented three different algorithms for computing the null
space of $\Jsq_i - \lambda I$ for medium and large blocks.
\begin{small}
\begin{enumerate} 
\item Randomized rank-revealing QR (RQR).  The algorithm performs
      two standard QR factorizations of dense matrices without pivoting.
      Although we do not take advantage of the sparsity of $\Jsq_i$
      in this approach, it is more efficient than other approaches when 
      the dimension of the desired null space is relatively large (e.g.
      larger than 10\% of the dimension of $\Jsq_i$).
\item Shift-invert Lanczos (SIL),
      which requires solutions of sparse linear systems. 
\item Polynomial accelerated subspace iteration (PASI).
We apply a standard subspace iteration \cite{sevbook} to the matrix
$p(\Jsq_i)$, where $p(\omega)$ is a polynomial that assumes the value of
1 at $\omega=\lambda$, and has a much smaller magnitude (than 1) in other
parts of the spectrum of $\Jsq_i$.
\end{enumerate}
\end{small}

Table~\ref{tab:dist_comparison} shows that our load balance scheme is 
much better than  a brute-force approach of distributing $\Jsq_i$ in a cyclic fashion
to different processors.  Table~\ref{tab:compare_methods} shows that 
PASI is more efficient when J=0.  The randomized QR algorithm appears to be
more efficient for larger J values.  However, when J becomes very large, which 
typically results in smaller dimension of the null space, PASI becomes
more efficient again.

\begin{small}
\begin{table}[ht]
\begin{minipage}[b]{0.5\linewidth}\centering
\renewcommand{\arraystretch}{1.1}
\caption{A comparison between the greedy load balancing algorithm with 
	        a parallel algorithm based on a cyclic distribution of $\Jsq_i$ blocks	
	        in terms of wall clock time (in seconds). ($n_p$ is the number of processors.)}
  \vspace{0.2cm}
  \label{tab:dist_comparison}
  \begin{tabular}{|c || c c c | c c |}
    \hline
    & & & & \multicolumn{2}{c|}{time (secs)} \\
    \cline{5-6}
    core & $\Nmax$ & alg & $n_p$ & cyclic & greedy \\
    \hline
    \hline
    $^6$Li & 12  & PASI  & 120 & 131   & 132 \\
    $^{12}$C & 4 & PASI & 120 & 6.1    & 5.2 \\
    $^{12}$C & 6 & PASI & 496 & 608    & 295 \\
    \hline
    \hline
    $^6$Li & 12  & RQR    & 120 & 233   & 193 \\
    $^{12}$C & 4 & RQR   & 120 & 18.7  & 17.0 \\
    $^{12}$C & 6 & RQR   & 496 & 1220  & 900 \\
    \hline
  \end{tabular}
\end{minipage}
\hspace{0.5cm}
\begin{minipage}[b]{0.45\linewidth}\centering
  \renewcommand{\arraystretch}{1.1}
  \caption{RQR decomposition vs PASI for different $J$ 
	  values. Both methods use the greedy load balancing technique.
          Times are in seconds.}
  \vspace{0.5cm}
  \label{tab:compare_methods}
  \begin{tabular}{| c || c c | c c |}
    \hline
    & & & \multicolumn{2}{c|}{time (secs)} \\
    \cline{4-5}
    core &$(N_{\mathrm{max}}$,$J$)  & $n_p$ & RQR   & PASI \\
    \hline
    \hline
    $^6$Li & (12,  0)     & 120 & 193  & 132  \\
    $^6$Li & (12,  1)     & 120 & 195  & 464  \\
    $^6$Li & (12,  12)   & 496 & 140  & 95   \\
    \hline
    $^{12}$C & (6, 0)    & 496 & 900   & 295   \\
    $^{12}$C & (6, 1)    & 496 & 890   & $>1,800$ \\
    $^{12}$C & (6, 12)  & 496 & 840  & 105 \\
    \hline
  \end{tabular}
\end{minipage}
\end{table}
\end{small}
\section{Optimization}
Optimization plays a central role in the building of next-generation nuclear energy functionals, including functionals based on density functional theory (DFT) and/or ab initio calculations. 
In the case of DFT-based functionals, for example, a primary computational bottleneck 
is determining parameter values 
so that the functional agrees with data on a set of observables such as binding energies, radii, and odd-even staggering \cite{scidac09}. Mathematically, we need to solve the optimization problem
\begin{equation}
\min_x\left\{f(x) = \sum_{i=1}^o \left(\frac{d_i-s(\theta_i;x)}{\sigma_i}\right)^2 : l_j\leq x_j \leq u_j, \, j=1,\ldots, n \right\},
\label{eq:chi2}
\end{equation}
where $n$ parameter values must be determined from a set of data of $o$ observables. Challenges in solving this problem include the computational expense of, and the noise resulting from, the iterative calculations performed when simulating the theoretical observable $s(\theta_i;x)$, and the fact that derivatives of some simulated observables with respect to certain parameters $x_j$ may not be available (or even exist) for use by an optimization algorithm.

\begin{figure}[bt!]
\begin{minipage}{.45\linewidth}
\includegraphics[width=.90\linewidth]{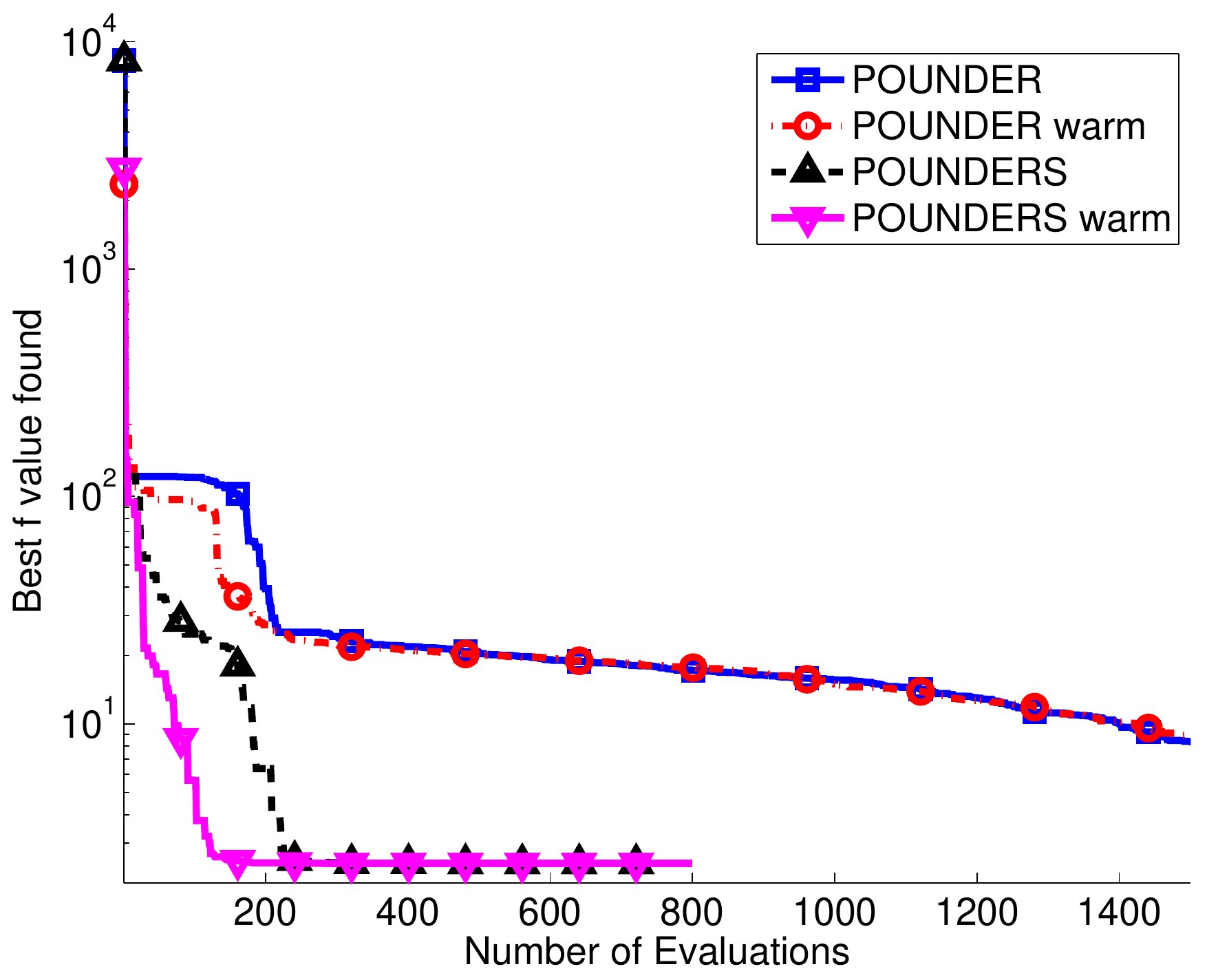}
\caption{\label{fig:pounder}Exploiting structure in parameter estimation problems substantially reduces the number of required simulation evaluations.}
\end{minipage}\hfill%
\begin{minipage}{.45\linewidth}
\includegraphics[width=.90\linewidth]{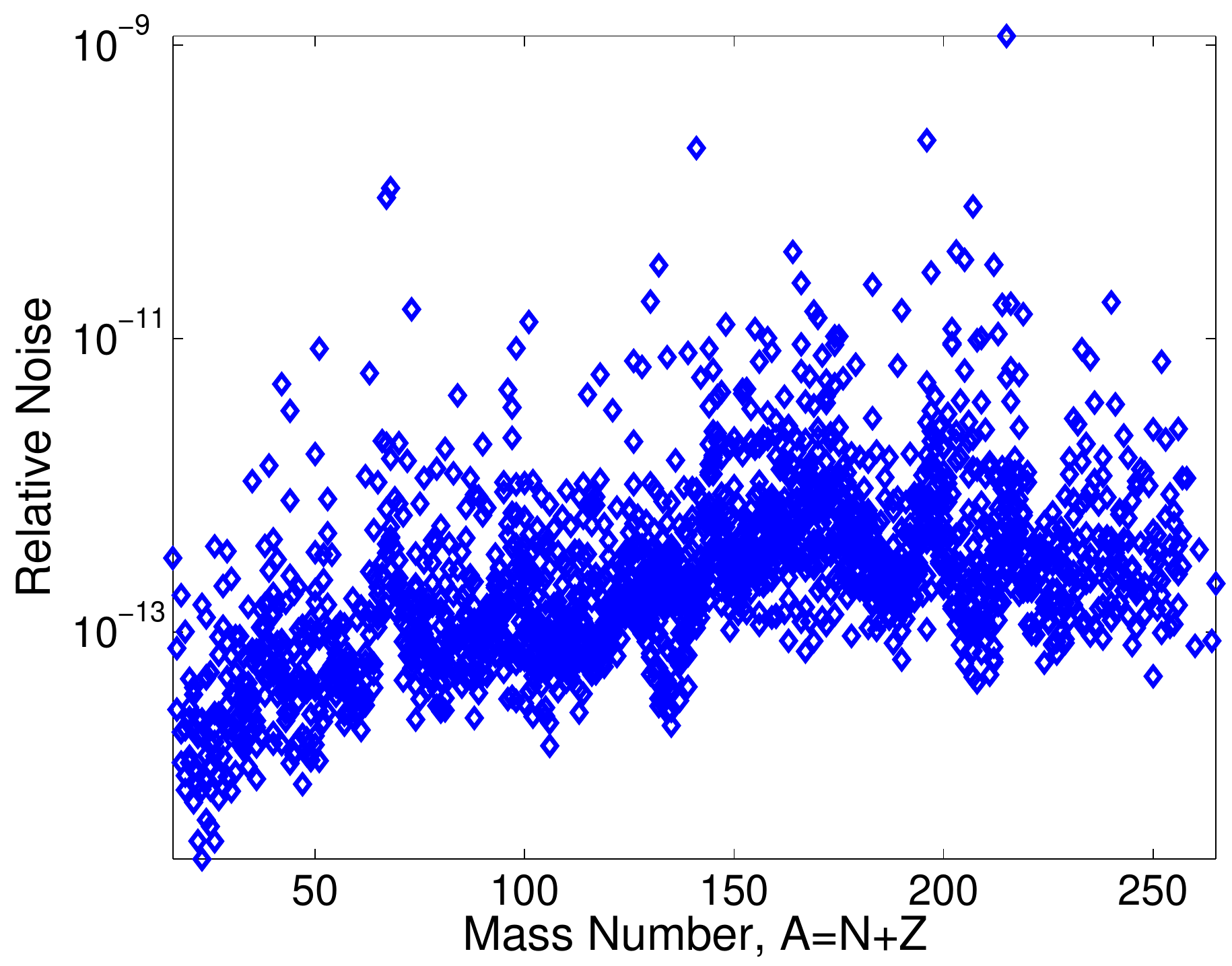}
\caption{\label{fig:noise}Quantifying the computational noise for a deterministic simulator across 2,049 nuclei.}
\end{minipage} 
\end{figure}

As part of SciDAC efforts, we have developed POUNDERS (Practical Optimization Using No DERivatives for Sums of squares), an algorithm for derivative-free optimization of nonlinear least-squares problems such as (\ref{eq:chi2}). A key benefit of POUNDERS is that it works with the individual residuals
${(d_i-s(\theta_i;x))}/{\sigma_i}$
rather than the aggregated fit function $f(x)$. As a result, POUNDERS can take advantage of the availability of the derivatives of some observables (e.g., binding energies) and can approximate nonlinearities in $f$ using simulations at fewer $x$ values.  As part of the TOPS collaboration, POUNDERS is now available through the open-source Toolkit for Advanced Optimization (TAO) \cite{TAO}.

Figure~\ref{fig:pounder} quantifies the computational savings in this ability to exploit the sums of squares structure in (\ref{eq:chi2}) for a fit to 2,049 binding energies. By working with the residuals, the POUNDERS variants obtain far better fits in far fewer evaluations than the analogous variants of POUNDER, a similar algorithm that does not have access to the residuals. The {\em warm\/} variants illustrate the benefit of using external simulations, done as part of an initial experimental design, to warm start the optimization. 

The savings in Figure~\ref{fig:pounder} can be substantial. For the more complex functional optimized  in \cite{nedf0}, each evaluation of $f$ requires 14.4 CPU hours. The resulting parameterization is then used to perform a simulation of nuclei across the
nuclear table in a calculation requiring 9,000 processors for more than half a day \cite{scidac09}. 

Mathematical work has also contributed to the sensitivity analysis of nuclear energy functionals. Though the simulations are typically deterministic, the aforementioned computational noise can obfuscate the number of reliable digits in computed functional properties. The ECNoise algorithm described in \cite{ecnoise} estimates a standard deviation-like quantity using only a few simulations. Figure~\ref{fig:noise} illustrates the relative noise in the computed binding energies for 2,049 nuclei with the parameterization obtained from the POUNDERS optimization in Figure~\ref{fig:pounder}. 
Estimates of the noise can, for example, reveal limitations on the predictability of computed functional observables and can enable stable approximations of the noisy derivatives needed for sensitivity analysis. 


\section{Conclusions}

Our work on eigenvalue calculations has made several impacts on
nuclear structure calculations.  Earlier collaborations with nuclear
physicists led to significant improvements to an eigensolver for
configuration interaction calculation, which was subsequently used
in predicting the properties of $^{14}$F before the isotope was
observed experimentally.  The work described in this paper 
takes configuration interaction calculation one step further.  It
enables our physics collaborators to efficiently compute energy
states of a nuclei with a prescribed total angular momentum instead
of computing many energy states and identifying those corresponding
to the prescribed total angular momentum.

The optimal parameters we have delivered to our physics
collaborators have resulted in realistic functionals that are
now being explored by a variety of groups outside of the UNEDF 
collaboration (as evidenced in the most recent JUSTIPEN conference
(\url{http://massexplorer.org/justipen/index.php})).
For example, our current results show remarkable power for predicting 
fission barrier heights, which is a first step toward a microscopic
understanding of fission. These results are a consequence of including a 
richer set of experimental data and more free parameters, resulting in 
problems that can be solved only by an efficient, state-of-the-art 
optimization algorithm.


\ack 
{\footnotesize
{\begingroup
The work under TOPS at ANL and LBNL
was supported by the Office of Advanced Scientific Computing Research of
the U.S. Department of Energy under contracts 
DE-AC02-06CH11357 (ANL), 
and
DE-AC02-05CH11231 (LBNL).
The UNEDF SciDAC collaboration was supported by
the U.S. Department of Energy under
grant numbers DE-FC02-09ER41582.
This work was also supported by
the Office of Nuclear Physics of the U.S.~Department of Energy
under grant numbers
DE-FG02-87ER40371 (Iowa State), 
DE-FG02-07ER41529 (Univ.~of Tennessee) 
and DE-FG02-96ER40963 (Univ.~of Tennessee). 



Computational resources were provided through an
INCITE award (James Vary, PI) at ORNL and ANL, and by the 
Laboratory Computing Resource Center (LCRC) at ANL, 
the National Energy Research Scientific Computing Center (NERSC) at LBNL,
and the National Center for Computational Sciences at ORNL. 
\endgroup}}

\end{document}